\documentclass[12pt]{iopart}
\usepackage{dcolumn}
\usepackage{bm}
\usepackage{graphicx}
\usepackage{xcolor}
\usepackage{graphics}
\usepackage{amssymb}
\begin{document}

\title[Element-specific visualization of dynamic magnetic coupling in a Co/Py bilayer microstruct.]{Element-specific visualization of dynamic magnetic coupling in a Co/Py bilayer microstructure}
\author{T. Feggeler$^{1,*}$, R. Meckenstock$^1$, D. Spoddig$^1$, C. Sch{\"o}ppner$^1$, B. Zingsem$^{1,2}$, T. Schaffers$^{3}$, H. Ohldag$^{4,**}$, H. Wende$^1$, M. Farle$^{1,5}$, A. Ney$^3$ and K. Ollefs$^1$}
\address{$^1$Faculty of Physics and Center for Nanointegration Duisburg-Essen (CENIDE), University of Duisburg-Essen, 47048 Duisburg, Germany}
\address{*Present address: Advanced Light Source, Lawrence Berkeley National Laboratory, Berkeley, CA, United States} 
\address{$^2$Ernst Ruska Centre for Microscopy and Spectroscopy with Electrons and Peter Gr\"unberg Institute, Forschungszentrum J\"ulich GmbH, 52425 J\"ulich, Germany}
\address{$^3$Institute of Semiconductor and Solid State Physics, Johannes Kepler University Linz, 4040 Linz, Austria} 
\address{$^4$SLAC National Accelerator Laboratory, 94025 Menlo Park, CA, United States} \address{Department of Physics, University of California Santa Cruz, Santa Cruz  CA 95064, United States}
\address{**Present address: Advanced Light Source, Lawrence Berkeley National Laboratory, Berkeley, CA, United States and Department of Material Sciences and Engineering, Stanford University, Stanford CA 94305, United States} 
\address{$^5$Kirensky Institute of Physics, Federal Research Center KSC SB RAS, Russia}
\ead{thomas.feggeler@uni-due.de}

\begin{abstract}
We present the element-specific and time resolved visualization of uniform ferromagnetic resonance excitations of a Permalloy (Py) disk - Cobalt (Co) stripe bilayer microstructure. The transverse high frequency component of the resonantly excited magnetization is sampled in the ps regime by a combination of ferromagnetic resonance (FMR) and scanning transmission X-ray microscopy (STXM-FMR) recording snapshots of the local magnetization precession of Py and Co with nanometer spatial resolution. The approach allows us to individually image the resonant dynamic response of each element, and we find that angular momentum is transferred from the Py disk to the Co stripe and vice versa at their respective resonances. The integral (cavity) FMR spectrum of our sample shows an unexpected additional third resonance. This resonance is observed in the STXM-FMR experiments as well and our microscopic findings suggest that it is governed by magnetic exchange between Py and Co, showing for the Co stripe a difference in relative phase of the magnetization due to stray field influence.
\end{abstract}
\ioptwocol
\maketitle

\section{Introduction}
For future information technology new concepts are needed involving the charge of the electron as well as its spin as information unit \cite{Wolf2001}. \textcolor{blue}{Several approaches for magnetism based logic have been introduced, ranging from soliton based concepts  \cite{Cow2000}, to magnonics in the form of e.g. genetically engineered magnonic computing \cite{Zing2019, Bar2021} to overcome the various limitations, e.g. thermal load and energy needs, encountered by modern computer technology.} This field of spintronics and magnonics requires to study even smaller magnetic structures in the gigahertz and terahertz regime.

Spin-based devices usually consist of more than one material, requiring the understanding of the element-specific dynamic magnetic properties and the resulting spin wave modes on the nanometer scale. X-ray detected ferromagnetic resonance (XFMR) \cite{BCK04,MKN11,BCK2013,GRW05,BRB05,GRW10,BRK09,BRB09,OMS15}, combining ferromagnetic resonance (FMR) with element specific magnetometry by means of X-ray Magnetic Circular Dichroism (XMCD) (see \cite{Dur09,vdLF04} and references therein) is a unique tool to address this challenge.

In this study Scanning Transmission X-ray Microscopy detected FMR (STXM-FMR) \cite{BKC15} has been used, offering temporal sampling down to 17~ps and nominal sub 50~nm lateral resolution in transverse XFMR geometry with a continuous wave excitation of the sample \cite{ChB12,BKC15,SMS17,GTF19,SSG19}. Uniform and non uniform resonant responses on the micro- \cite{Pile2020a,Pile2020b} and sub 50~nm nanometer \cite{Feggeler2021} scale have been monitored and analysed. Here we investigate resonant excitations of a bilayered microstructure consisting of a Cobalt (Co) stripe deposited on a Permalloy (Py) disk with element specificity. Earlier studies of equally dimensioned ultra-thin ferromagnetic bilayers (thickness usually about or below 10~nm) showed two uniform resonance modes, typically explained as an in-phase and out-of-phase optical or acoustical modes e.g. \cite{HPD1988}. In the conventional FMR measurements of our bilayer microstructure with a total thickness of 60~nm the individual resonances of the Py and Co microstructures are identified.  \textcolor{blue}{In addition, a third resonance in both materials is seen, which cannot be explained by the aforementioned approach for equally dimensioned ultra-thin bilayers, but by Py and Co resonating in phase as an entity, mediated by exchange coupling. Thus, by our spatially, time and element specific STXM-FMR the origin of the three resonances is revealed, visualizing as well local phase and amplitude variations, which are not visible in conventional FMR spectra} 
 

\section{Experimental details}
We measure FMR excitations in their linear regime using a micro-resonator based, element-specific and spatially resolved STXM-FMR setup realized at Stanford Synchrotron Radiation Lightsource (SSRL)\cite{BKC15,SMS17}. The sample is a polycrystalline Co stripe (2.0~$\mu$m~length, 0.5~$\mu$m~width, 30~nm~thickness) deposited on a polycrystalline Permalloy(Py)-disk with 2.5~$\mu$m diameter and 30 nm thickness (see Fig. 1a)). It is fabricated by a three step lithography and electron-beam deposition of the ferromagnetic material \cite{Posth2009} on a 200 nm thick Si\textsubscript{3}N\textsubscript{4} membrane. To measure the FMR spectrum the sample is positioned in the omega-shaped loop of a micro-resonator offering a sensitivity of 10\textsuperscript{6} $\mu$\textsubscript{B} \cite{BNH2011, Mec08, SWS2014}. The sample is excited by a homogeneous linearly polarized microwave field with an amplitude of $\leq$ 1.5~mT. A STXM image of the sample using a step size of 100 nm is shown in Fig. 1b).

Fig. 1c) pictures the conventional FMR spectrum of the sample obtained at a microwave frequency of 9.27~GHz in a magnetic field B\textsubscript{Ext}=0~-~200~mT featuring three main resonances, the first resonance at B\textsubscript{Ext,1}=58.3~mT, the second resonance  at B\textsubscript{Ext,2}=84.9 mT, and the third resonance at B\textsubscript{Ext,3}=112.7 mT. Resonances 1 and 3 are modes of the Py disk and the Co stripe, respectively. Magneto-crystalline anisotropy is neglible in both samples due to their polycrystallinity. The Co stripe exhibits the highest resonance field due to shape anisotropy despite its high M\textsubscript{sat}~=~1420~kA/m \cite{Jiles2016} considering the stripe geometry and the perpendicular orientation of its long side to B\textsubscript{Ext} (Fig. 1a), while Py with M\textsubscript{sat}~=~860~kA/m \cite{Jiles2016} shows the lowest resonance field. This is confirmed by angular dependent FMR measurements of a Co stripe \cite{SWS2014}. The origin of the intermediate resonance 2, however, can not readily be understood, because one would only expect the two individual resonances. The presence of a third resonance at an intermediate field strongly suggests that Co and Py resonate as one entity resulting in a coupled uniform resonance. Although it appears reasonable to assume that this is due to exchange coupling across the interface, it is not possible to directly deduce \textcolor{blue}{the} microscopic mechanism behind our observation from the classical FMR spectrum. To elucidate this, we use STXM-FMR. 


\begin{figure}[!tb]
\resizebox{1\columnwidth}{!}{\includegraphics{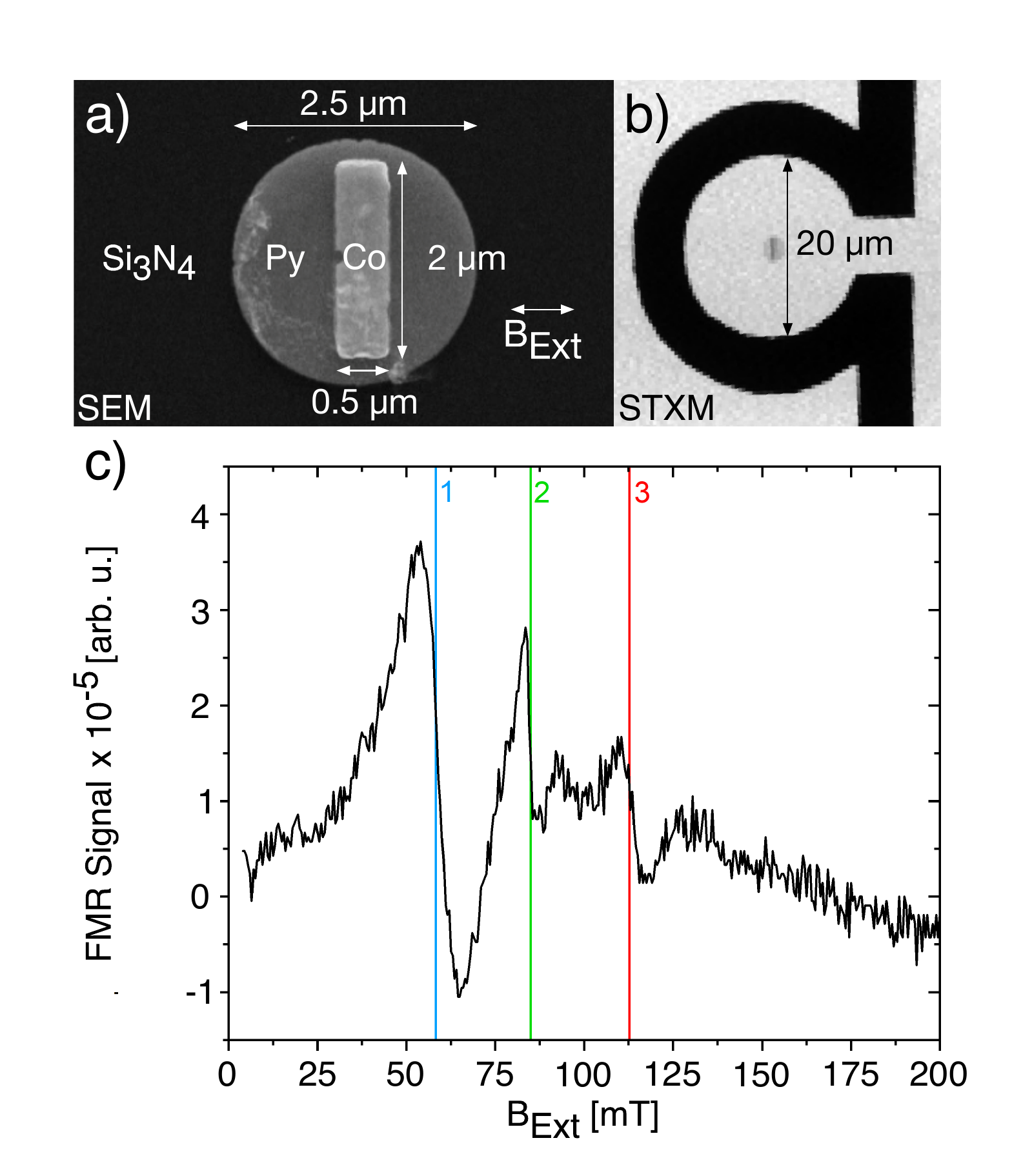}} 
\caption{a) Scanning electron microscopy (SEM) image of the Co stripe/Py disk bilayer on a Si\textsubscript{3}N\textsubscript{4} membrane. The orientation of B\textsubscript{ext} is indicated; b) STXM image of the sample in the micro-resonator loop; The high-frequency magnetic field oscillates in the out-of-plane direction c) FMR spectrum of the sample shown in a) with four major resonances, 1: Py resonance, 2: Coupled resonance, 3: Co stripe center resonance, 4: Resonance of the long sides of the Co stripe.\label{fig1}}
\end{figure}

In STXM-FMR the sample is mapped by a focussed X-ray beam (energy tunable between 200~eV and 1200~eV at the SSRL), while the transmitted intensity is detected by an avalanche X-ray photodiode. B\textsubscript{ext} is applied in the sample plane along the short axis of the Co stripe (Fig.~1a)) with perpendicular orientation to the incident circularly polarized X-rays. The time-dependent transverse component of the magnetization at 9.129~GHz is probed by means of the XMCD effect, for details see \cite{BKC15}. The magnetization oscillation is sampled with 6 consecutive images separated by a static phase difference of 60$^\circ$ (18~ps), each with and without applied microwave excitation. To extract the microwave induced X-ray absorption the respective difference of both datasets is taken. Fig.~2a) shows the resulting 6 STXM-FMR images at the Co L\textsubscript{3}-edge with an applied external magnetic field of 112.7 mT (Fig.~1b)). Brighter and darker contrast indicates a lower/higher X-ray absorption than the average. The contrast within the area of the Co stripe indicates a microwave induced response. Thus, the bright and dark contrast in Fig.~2a) shows the deviations of the magnetization from its equilibrium orientation along the oscillation axis of the high frequency magnetic field. Fig.~2b) shows the oscillation of the STXM-FMR signal at the position of the Co stripe. Its maximum is visible at a relative phase of about 90$^\circ$. The black curve in Fig.~2b) was recorded at an off resonance field of 30~mT and thus the Co is only driven by the microwave field. The red STXM-FMR signal is shifted by 90$^\circ$, as generally expected for a resonant response \cite{Pain2005}.

\begin{figure}[!tbp]
\resizebox{1\columnwidth}{!}{\includegraphics{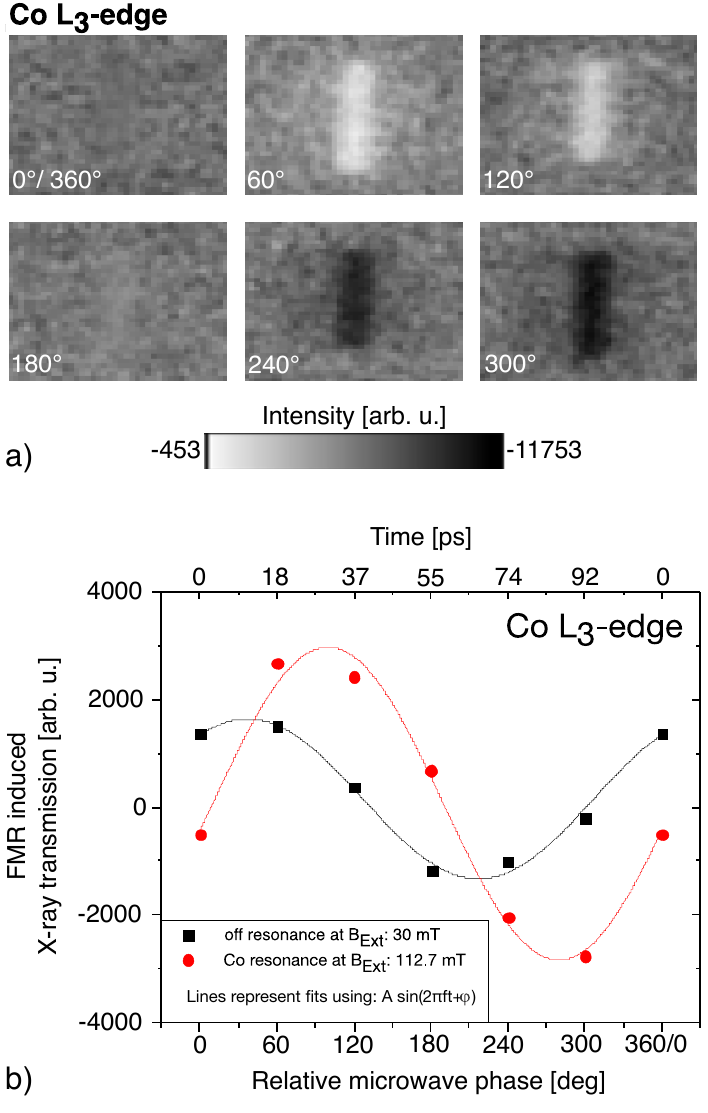}}
\caption{a) 6 STXM-FMR difference images obtained from six microwave on and six microwave off images recorded every 60$^\circ$ (18~ps) at the Co  L\textsubscript{3}-edge at B\textsubscript{Ext,3}. b) FMR induced X-ray transmission signal (red dots: at Co resonance 3, black squares: off resonance) as function of time.\label{fig2}}
\end{figure}

\begin{figure}[!ht]
\resizebox{1\columnwidth}{!}{\includegraphics{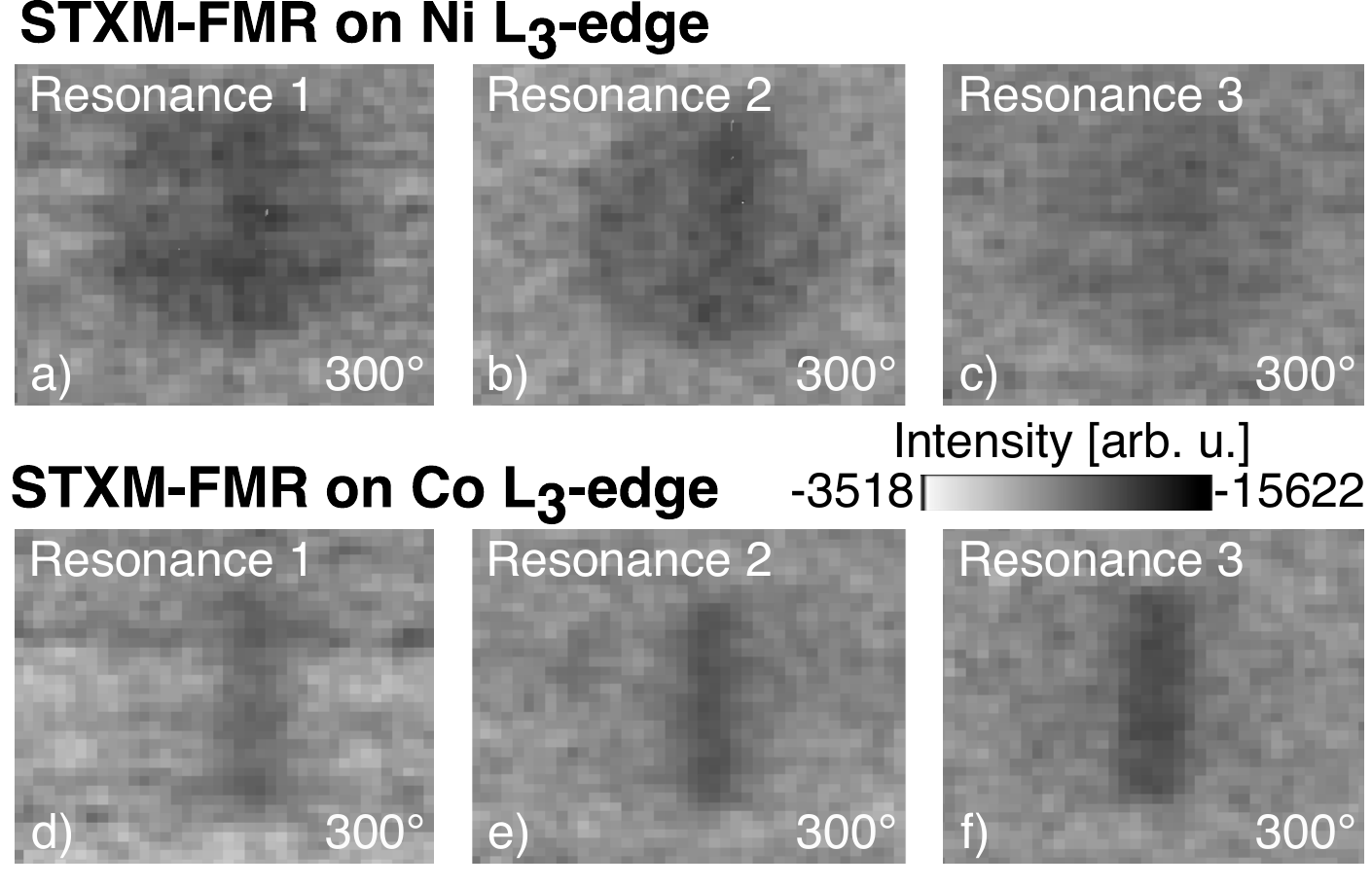}}
        \caption{STXM-FMR images taken at the respective B\textsubscript{ext} for resonance 1, 2, and 3. a)-c) are recorded at the Ni  L\textsubscript{3}-edge, d)-e) are recorded at the Co L\textsubscript{3}-edge. The STXM-FMR images correspond to the time slot at 92~ps (300$^\circ$).\label{fig3}}
\end{figure}

\section{Results and discussion}

The element-specific and spatially resolved measurements depicted in Fig. 3 show the STXM-FMR images at the Ni L\textsubscript{3}-edge (852~eV) (Fig.~3a)-c)), and Co L\textsubscript{3}-edge (779~eV) (Fig.~3d)-f)) taken at a relative phase of 300$^\circ$ exhibiting the highest contrast, with a 100~nm step size and a dwell time of 5000~ms. Grey contrast corresponds to an average contrast value, which is set to the same background color level for all images of the figure. In Fig. 3a)-3c) the complete spherical area of the Py disk shows STXM-FMR contrast at all three resonance fields, indicating a resonant response of the Py disk. Each of STXM-FMR images in Fig. 3a)-3c) reveals a darker colored contrast area at the location of the on-top lying Co stripe. Fig. 3d)-f) picture the STXM-FMR contrast originating from Co stripe while the Py disk is almost invisible with the The STXM-FMR image at resonance 3 (Fig. 3f)) showing the darkest coloured contrast of all the images (Fig. 3d)- f)).

The STXM-FMR image taken at resonance 1 shown in Fig.~3a) (Ni L\textsubscript{3}-edge) shows a uniform contrast distribution within the disk area with a higher intensity contrast area at the position of the Co stripe. This is corresponding to a homogeneous uniform resonant response of the Py disk as expected from the conventional FMR spectrum (Fig. 1a)). The contrast visible in Fig.~3d) within the Co stripe originates from Co being driven by the Py in resonance, inducing a slight increased precessional motion in the Py disk (higher intensity contrast area in Fig. 3a)) mediated by exchange coupling. The corresponding observation is made for the Py disk at B\textsubscript{ext,3}, \textcolor{blue}{where the Py magnetic moments are less agile due to their alignment along the direction of B\textsubscript{ext}, resulting in the Py disk getting only slightly driven by the Co in resonance.} In consequence the contrast of the driven Co in Fig.~3d) is more intense than the one of the driven Py in Fig.~3c) as the Co moments are not completely aligned along B\textsubscript{ext,1} and therefore are more agile and easily driven compared to the Py moments of the disk at B\textsubscript{ext,3}. This excitation between the two constituents across the interface illustrates a transfer of angular momentum (spin current) between the two magnetic materials due to exchange coupling.

At resonance 2 the STXM-FMR images at both absorption edges show uniformly distributed contrast at the location of both of the sample constituents, indicating a coupled resonance originating from exchange coupling between Py and Co both being in resonance and contributing to the STXM-FMR signal. \textcolor{blue}{Such modes have been observed before in multilayer films and originate in the interface exchange between both constituents. The exchange length in Co and Py is several nanometers, thus in ferromagnetic resonance the sample behaves in this area alloy-like. This can be seen for example in \cite{Kordecki1991}, where spin wave spectra exist at an effective magnetization of FeNi as an alloy-like entity in addition to the individual spin-wave resonance of Fe. Our STXM-FMR measurements of resonance 2 show at the Ni L\textsubscript{3} edge that the resonance can be observed in the whole area of the disc with a darker contrast at the position of the Co stripe/interface, broader than the contrast seen at the Co L\textsubscript{3} edge, since the edge spins of Co are still not aligned along B\textsubscript{ext,2} and thus not in resonance. This is due to that beside the exchange length for this excitation the coherence length of the FMR precession is important, which ranges depending on the material up to several millimeters (e.g. 7 mm for YIG \cite{Pelzl1994}). In consequence the intensities of the three resonance modes shown in Fig. 2 b) consistent to this interpretation. The Py resonance 1 shows the highest intensity due to the largest sample volume, therefore the coupling resonance 2 shows less intensity as it originates as described above only from a part of the sample, while the Co resonance 3 exhibits the lowest intensity, corresponding to the smallest excited volume.}

An amplitude and phase analysis of the recorded 6 STXM-FMR images \cite{Zingsem2019} at resonance 2, shown in Fig. 4a)-b) at the Ni and Co L\textsubscript{3}-edge,  reveals further details on the origin of the FMR excitations, not directly visible in the grayscale plots. After normalising the STXM-FMR data to the average intensity of each image, a sine fit is applied to the time evolution of each pixel. Thus, the pixels of the STXM-FMR image can be color coded representing amplitude, phase and fit accuracy obtained from the sinusoidal fits. The color coding was chosen as such that bright pixels represent \textcolor{blue}{a large amplitude}, the phase is represented as the hue value and pixels with very high saturation indicate a high fit accuracy \textcolor{blue}{by encoding the p-value obtained from the fit as color saturation.} Thus, Fig. 4a) indicates a homogeneously distributed relative phase of approximately 90 degrees to the exciting microwave inside the whole Py disk, whereas the amplitude of the Py excitation is largest at the position of the upper and lower edge of the Co stripe, which is not directly visible in Fig.~3b). Fig.~4b) shows bright and saturated colored pixels only at the position of the Co stripe, depicting different phase values between the center (approx. 90~degrees as with the Py) and the upper and lower edges (poles) of the Co stripe (approx. 60~degrees), \textcolor{blue}{due to, as for a typical bar magnet, the stray field influence.} This leads to a different phase at the top and bottom edges of the Co stripe. The local phase change is only resolvable with our technique. Fig. 4a)-b) prove that the exchange coupled resonance 2 is mainly excited at the directly overlapping areas of disk and stripe and is not due to either an optical or acoustical mode excitation, \textcolor{blue}{Both Py and Co are resonating at the field value corresponding to the one of an alloy-like entity, which is proven by similar phases.}

The STXM-FMR observation of the three resonances is in agreement with the line widths observed in Fig. 1c. The largest peak to peak line width of 15~mT is observed at resonance 1, where the whole Py disk is in resonance but drives the Co moments at the area of the Co stripe. This yields an additional damping for the Py and an additional line distribution resulting from the area outside and below the stripe. The same is valid for the Co resonance 3 (peak to peak line width of 10~mT), there the Co is driving the Py underneath, which is already completely aligned with the external field and thus provides a stronger damping. The coupling resonance 2 has the smallest peak to peak line width (about 5~mT). In addition, an asymmetric line shape of the resonance is visible indicating a distribution of different excitations.

\begin{figure}[!tb]
\resizebox{1\columnwidth}{!}{\includegraphics{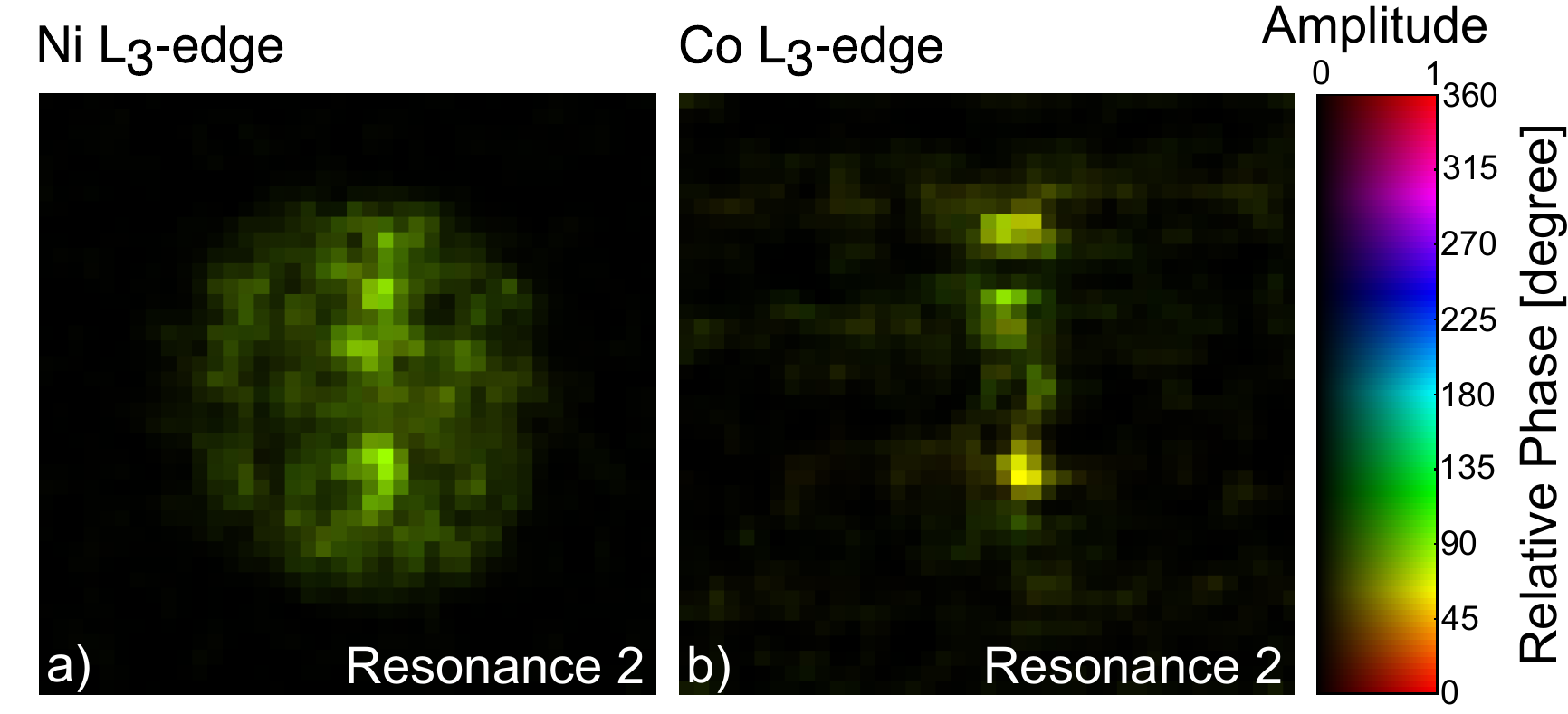}}
\caption{a), b) Result of the pixel-wise fit analysis of the STXM-FMR images, the color coding of the amplitude and relative phase is displayed by the color bar. All images represent resonance 2.\label{fig4}}
\end{figure}

\section{Conclusion}
The magnetization dynamics of a coupled Py disk Co stripe bilayer microstructure has been analyzed in the linear response regime with element specificity, spatial, time and phase resolution. At the Py resonance the Co magnetization is driven into precession by angular momentum transfer mediated by exchange coupling of the precessing Py.  We show in our experiment that a coherently precessing spin polarization is transferred via inter-material exchange at the interface to the ferromagnetic material, which is not in resonance.

In earlier investigations of extended ferromagnetic ultra-thin bilayers two main resonances have been observed attributed to an in-phase and out-of-phase optical or acoustical mode \cite{HPD1988}. In contrast here we revealed in the bilayered microstructure the occurrence of a third main resonance mode, which is explained by Py and Co resonating as an exchanged coupled entity. Using an amplitude and phase analysis method an inhomogeneous excitation of the Co stripe at the coupled resonance is revealed, due to the stray field effects at the poles of the stripe, whose influence is visualized by micromagnetic simulations. Thus, this mode is identified as an exchange coupled and dipolarly influenced excitation of the Co/Py disk stripe microstructure.

\section{Acknowledgement}
The authors would like to thank the German Research Foundation (DFG project: 321560838 (OL513/1-1)) and the Austrian Science Fund (FWF project: I 3050-N36) for financial support.
We gratefully acknowledge the experimental assistance of S. Bonetti during beamline setup.
The use of the Stanford Synchrotron Radiation Lightsource, SLAC National Accelerator Laboratory, is supported by the U.S. Department of Energy, Office of Science, Office of Basic Energy Sciences under Contract No. DE-AC02-76SF00515.

\end{document}